\documentclass[12pt,fleqn]{article}
\usepackage{amsmath}

\begin{document}

\title{\textbf{Transformation of a generalized Harry Dym
equation into the Hirota--Satsuma system}}

\author{\textsc{S.~Yu.~Sakovich\footnote{Home institution:
Institute of Physics, National Academy of Sciences, 220072
Minsk, Republic of Belarus. Contact e-mail: saks@pisem.net}}%
\bigskip\\{\footnotesize Mathematical Institute, Silesian
University, 74601 Opava, Czech Republic}}

\date{}

\maketitle

\begin{abstract}
The new generalized Harry Dym equation, recently introduced by
Z.~Popowicz in Phys.\ Lett.\ A 317, 260--264 (2003), is
transformed into the Hirota--Satsuma system of coupled KdV
equations.
\end{abstract}

\section{Introduction}

Recently, Popowicz \cite{Pop} introduced the following new
generalization of the Harry Dym (HD) equation:
\begin{equation}
\label{e1}
u_t = u^3 \left( u^{-1/2} v^{3/2} \right)_{xxx}, \qquad
v_t = v^3 \left( v^{-1/2} u^{3/2} \right)_{xxx}.
\end{equation}
This remarkably simple and symmetric system of coupled
equations possesses a Lax pair and a Hamiltonian structure
\cite{Pop}.

In the present paper, we construct a chain of
transformations which relates the new generalized HD
equation of Popowicz \eqref{e1} with the well-studied
Hirota--Satsuma system of coupled KdV equations \cite{HS}.
There are no general methods of transforming a given nonlinear
system into another one, less complicated or better studied,
and the usual way of finding necessary transformations is based
on experience, guess and good luck. For this reason, we give no
comments on how these transformations were found in the present
case.

\section{Transforming the generalized HD equation}

First, we transform the dependent variables $u$ and $v$ of the
generalized HD equation \eqref{e1} in the following way:
\begin{equation}
\label{e2}
u = a \exp (b), \qquad v = a \exp (-b).
\end{equation}
In the new dependent variables, $a(x,t)$ and $b(x,t)$, the
system \eqref{e1} takes the form
\begin{equation}
\label{e3}
\begin{split}
a_t &= a^3 a_{xxx} + 12 a^3 a_x b_x^2 + 12 a^4 b_x b_{xx},\\
b_t &= - 2 a^3 b_{xxx} - 6 a^2 a_x b_{xx} - 6 a^2 a_{xx} b_x
- 8 a^3 b_x^3,
\end{split}
\end{equation}
where the (original) HD equation is clearly seen at
$b = \mathrm{constant}$. Note that the separants of the coupled
evolution equations \eqref{e3}, i.e.\ the coefficients at the
third-order $x$-derivatives in \eqref{e3}, have the form
$\mathrm{constant} \times a^3$, and thus the role played by the
invertible transformation \eqref{e2} is to prepare the system
\eqref{e1} for the next step of transforming, which brings
\eqref{e3} into a constant-separant form.

Second, we try to transform $x$, $a$ and $b$ in \eqref{e3} as
follows:
\begin{equation}
\label{e4}
x = p(y,t), \qquad a(x,t) = p_y(y,t), \qquad b(x,t) = q(y,t).
\end{equation}
This is an extension of the transformation used by Ibragimov
\cite{Ibr} to relate the (original) HD equation with the
Schwarzian-modified KdV equation. The transformation \eqref{e4}
really works and relates the system \eqref{e3} with the
constant-separant system
\begin{subequations}
\label{e5}
\begin{align}
p_t &= p_{yyy} - \frac{3 p_{yy}^2}{2 p_y} + 6 p_y q_y^2,
\label{e5a} \\
q_t &= - 2 q_{yyy} - 2 q_y^3 - 3 \left( \frac{p_{yyy}}{p_y}
- \frac{3 p_{yy}^2}{2 p_y^2} \right) q_y.
\label{e5b}
\end{align}
\end{subequations}
To verify this, one may use the identities
\begin{equation*}
a \partial_x = \partial_y, \qquad
a_t = p_{yt} - \frac{p_{yy} p_t}{p_y}, \qquad
b_t = q_t - \frac{q_y p_t}{p_y}
\end{equation*}
which follow from \eqref{e4} straightforwardly. Note that
\eqref{e4} is not an invertible transformation: it maps the
system \eqref{e5} into the system \eqref{e3}, whereas its
application in the opposite direction, from \eqref{e3} to
\eqref{e5}, requires one integration by $x$. We have omitted
the terms $\alpha (t) p_y$ and $\alpha (t) q_y$ in the
right-hand sides of \eqref{e5a} and \eqref{e5b},
respectively, where this arbitrary function $\alpha (t)$
appeared as a `constant' of that integration. The
Schwarzian-modified KdV equation is clearly seen in the system
\eqref{e5} at $q = \mathrm{constant}$. Nevertheless, further
steps of transforming the generalized HD equation \eqref{e1}
do not involve the Schwarzian derivative.

Third, we make the transformation
\begin{equation}
\label{e6}
f(y,t) = \frac{p_{yy}}{p_y}, \qquad g(y,t) = q_y,
\end{equation}
admitted by the equations \eqref{e5} owing to their form, and
obtain the system
\begin{equation}
\label{e7}
\begin{split}
f_t &= \left( f_{yy} + 12 g g_y - \tfrac{1}{2} f^3
+ 6 f g^2 \right)_y,\\
g_t &= \left( - 2 g_{yy} - 3 g f_y + \tfrac{3}{2} f^2 g
- 2 g^3 \right)_y
\end{split}
\end{equation}
which belongs to the well-studied class of coupled mKdV
equations.

Fourth, we apply the Miura-type transformation
\begin{equation}
\label{e8}
r(y,t) = f_y + \tfrac{1}{2} f^2 + 2 g^2, \qquad
s(y,t) = g_y + f g
\end{equation}
to the coupled mKdV equations \eqref{e7} and obtain the
Hirota--Satsuma system of coupled KdV equations \cite{HS}
\begin{equation}
\label{e9}
r_t = r_{yyy} - 3 r r_y + 24 s s_y, \qquad
s_t = - 2 s_{yyy} + 3 r s_y.
\end{equation}
Note that the system \eqref{e7} and the transformation
\eqref{e8} (up to a linear change of variables) were
introduced in \cite{Wil}.

Consequently, we have constructed the chain of transformations
\eqref{e2}, \eqref{e4}, \eqref{e6} and \eqref{e8} which
relates the generalized HD equation \eqref{e1} with the
Hirota--Satsuma system \eqref{e9}.

\section{Conclusion}

In this paper, we transformed the new generalized HD
equation of Popowicz \eqref{e1} into the well-known
Hirota--Satsuma system of coupled KdV equations \eqref{e9}.
Possible applications of this, which were completely out of the
scope of our work, may include relating Lax pairs and
Hamiltonian structures of \eqref{e1} and \eqref{e9}, deriving a
recursion operator for \eqref{e1} from the known one of
\eqref{e9}, explaining analytic properties of \eqref{e1} from
the standpoint of the Painlev\'e property of \eqref{e9}, etc. We
can add that chains of transformations, similar to the one
obtained in this paper, have been successfully used in
\cite{Sak} for relating two new generalized HD equations of
Brunelli, Das and Popowicz \cite{BDP} with known integrable
systems as well.

\section*{Acknowledgments}

The author is deeply grateful to all people at the Mathematical
Institute of Silesian University for kind help and hospitality,
and to the Centre for Higher Education Studies of Czech
Republic for support.

\end{document}